# Full Realization Scheme of the Tensor Product Space of $N$ Distinguishable Photons in Two States

Avi Marchewka

8 Galei Tchelet St., Herzliya, Israel

Abstract

The ability to control and hence to realize a given number of photons is of major interest from a fundamental point of view. e.g. Bell inequalities, photons bunching. In recent years this interest has grown by the so called the "Second Quantum Revolution" where such an ability is needed for quantum computers, etc. In this paper we show that such a realization can't be done by a unitary process. Therefore, a non-unitary interferometer is given to build a full realization of the tensor product space for two photons at two states. Finally, by modifying the previous interferometer, the full tonsorial product space of $N$ photons in two states is shown.

1. Introduction

In physics, it is a standard procedure to build a formal theory from an experiment's results. The theory gives a broader description of the concrete results at hand. The detailed realization of this wider description given by the theory may be a challenge by itself and often bear on technological uses.

Indeed, the construction of a single and a few photons has been one of the important landmarks in studying the fundamental aspects of quantum mechanics, e.g. the Bell inequality, photons bunching and so on [1]. Furthermore, nowadays one of the bases of the so-called the "second quantum revolution" is the ability to manipulate small numbers of degree of freedom, in particular a few photons, and hence the interest to develop techniques to produce and control several photons[2].

In this paper, we focus on giving a way to realize the general states of $N \geq 1$ distinguishable photons in two states, where the photons' frequencies will be used to distinguish between the photons. Such a realization seems to be new.

The description of $N$ distinguishable photons each in two states is given by the tensorial product of the photons states [3]

$$\left|1_{\omega^{(1)}}\right\rangle_{q_1,q_2} \otimes \left|1_{\omega^{(2)}}\right\rangle_{q_1,q_2} \cdots \otimes \left|1_{\omega^{(N)}}\right\rangle_{q_1,q_2}$$

Throughout the paper we use the notation that denotes the photon in given states in the lower number of its ket $\left|\ \right\rangle_1$ i.e., the photon in state one [4].

Whereas using a beam splitter and phase shifts to construct the single-photon realization, $N=1$, in two states, i.e. $\left|\chi\right\rangle = \alpha_1\left|1\right\rangle_1 + \alpha_2\left|1\right\rangle_2$ is a straightforward process, other cases are not obvious. Indeed, the realization of the product consisting of two photons in two states poses a challenge and requires some effort to overcome. As will be shown in the next sections, the difficulty derives from the fact that it is impossible to construct two photons (or more) in two states via unitary transformation.

However, it turns out that combining two single-photons each in single state via non-unitary transformation can be used to build a realization of two photons in two states. The type of non-unitary transformation we are going to use is the so-called post-selected measurements [4].

In section 2, we explain the aim of the paper in more detail. In section 3, we show that unitary transformation cannot be used to construct realization of the tensor product space of two photons in two states. In section 4, the non-unitary interferometer is given and used to construct the realization of two distinguishable photons in two states. In section 5, a modification of the previous interferometer is given in which the generalization of $N \geq 2$ distinguishable photons in two states is shown. In section 6, a summary and discussion are provided.

## 2. Aim and motivation

To set the ground of the paper's aim let us first consider a fictitious scenario of $N$ distinguishable particles denoted by the quantum number $|S_k\rangle$ with $k = 1, 2...N$. Additionally, consider that each of the $N$ particles is in a superposition of two states, say "energy" states $|E_1\rangle$ and $|E_2\rangle$,

$$|k\rangle = \alpha_{k,1}|S_k, E_1\rangle + \alpha_{k,2}|S_k, E_2\rangle \tag{1.1}$$

Where $\{\alpha_{k,1}, \alpha_{k,2}\} \in \mathbb{C}$ and due to normalization of the states $|\alpha_{k,1}|^2 + |\alpha_{k,2}|^2 = 1 \ \forall k$. Assuming that we can prepare each of the particles in its own superposition (1.1), the product states of the $N$ particles becomes,

$$|\psi\rangle = \prod_{k=1}^{N} \otimes |k\rangle \tag{1.2}$$

Then the question is how to realize the produced states (1.2) in the lab.

In the optics case, the above fictitious scenario can be recognized now as non-fictitious in the following manner: the $N$ distinguishable particles is now $N$ distinguishable photons with different frequencies, denoted by $|S_k\rangle$, and the two "energy" states are placing each photon in two arms of beam splitter, denoted by $|\ \rangle_1$ and $|\ \rangle_2$ Assuming that each of the photons with defined frequency can be put in superposition of the two arms,

$$|1_k\rangle = \alpha_{k,1}|S_k\rangle_1 + \alpha_{k,2}|S_k\rangle_2 \tag{1.3}$$

then the product states of the $N$ distinguishable photons in the two beam splitter arms becomes

$$|N\rangle = \prod_{k=1}^{N} \otimes |1_k\rangle \quad (1.4)$$

Now the aim of this paper is to show how to produce these product states.

3. The failure of unitary transformation in the realization of the product states (1.4) for $N \geq 1$ distinguishable photons

First, we want to show that it is impossible to realize two dimensional general product states by using unitary transformation. The general states of a single photon in two dimensional states space is given by

$$\left|1_{\omega^{(i)}}\right\rangle = \alpha_1^{(i)} \left|1_{\omega^{(i)}}\right\rangle_2 + \alpha_2^{(i)} \left|1_{\omega^{(i)}}\right\rangle_3 \quad (1.5)$$

Where $\{\alpha_1^{(i)}, \alpha_2^{(i)}\} \in \mathbb{C}$ are two normalized complex numbers i.e. $\left|\alpha_1^{(i)}\right|^2 + \left|\alpha_2^{(i)}\right|^2 = 1$, $\omega^{(i)}$ is the frequency of photon $i$, $i = 1, 2 \cdots N$.

The state of a single photon (1.5) has three parameters that define it, two phases and by normalization one amplitude. Furthermore, since a global phase has no physical significance, the number of the parameters of the state (1.5) is reduced to two parameters.

The product space of two distinguishable photons $\left|1_{\omega^{(1)}}\right\rangle$ and $\left|1_{\omega^{(2)}}\right\rangle$ $\omega^{(1)} \neq \omega^{(2)}$ each in two states is

$$\left|1_{\omega^{(1)}}, 1_{\omega^{(2)}}\right\rangle = \left(\alpha_1^{(1)} \left|1_{\omega^{(1)}}\right\rangle_1 + \alpha_2^{(1)} \left|1_{\omega^{(1)}}\right\rangle_2\right) \otimes \left(\alpha_1^{(2)} \left|1_{\omega^{(2)}}\right\rangle_1 + \alpha_2^{(2)} \left|1_{\omega^{(2)}}\right\rangle_2\right) \quad (1.6)$$

It follows that there are six parameters that define those product state. Again, using the physical insignificance of the global phase reduces the number of parameters of the state (1.6) into four.

The beam splitter may be represented by a two dimensional matrix

$$e^{i\Theta} \begin{pmatrix} \cos\theta & i\sin\theta \\ i\sin\theta & \cos\theta \end{pmatrix} \quad (1.7)$$

It is convenient to set the global phase of the beam splitter to $\Theta = 0$, then by (1.7) the beam splitter has one degree of freedom. It follows that it is impossible to construct a general state of one photon by a single beam splitter.
However, we can add a phase shifter to each of the beam splitter output arms (see fig 1), thus

$$u(\varphi_1,\varphi_1,\theta) = \begin{pmatrix} e^{i\varphi_1} & 0 \\ 0 & e^{i\varphi_2} \end{pmatrix} \begin{pmatrix} \cos\theta & i\sin\theta \\ i\sin\theta & \cos\theta \end{pmatrix} = \begin{pmatrix} e^{i\varphi_1}\cos\theta & ie^{i\varphi_1}\sin\theta \\ ie^{i\varphi_2}\sin\theta & e^{i\varphi_2}\cos\theta \end{pmatrix} \qquad (1.8)$$

Now we have additional two degrees of freedom and together with the beam splitter three degrees of freedom in total. Then the state (1.5) can be realized.

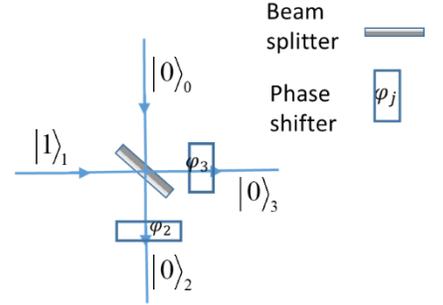

In figure 1 a beam splitter and two phase shifters are used to realize the states (1.5). This can be done by setting the beam splitter transmit and reflect coefficients $t$ and $r$ respectively and the two phase shifters $\varphi_2, \varphi_3$ as follows,

$$|\alpha_1| = |r| \quad , \quad \arg(\alpha_1) = -\frac{\pi}{2} + \varphi_2$$
$$|\alpha_2| = |t| \quad , \quad \arg(\alpha_2) = \varphi_3$$

Figure 1: a realized a single photon in two states eq. (1.5) via a beam splitter and two phase shifters.

In the same way, since the product states (1.6) has four parameters the system in fig. 1 cannot be used to realize it. One may wonder if by adding additional beam splitters and phase shifters the realization of the states (1.6) may be possible. To see why the answer to this question is negative consider $k$ systems constructed from beam splitters and phase shifters, each of them given by the transformation (1.8). Then, the overall transformation has the following isomorphic relation,

$$U(\theta_1,\varphi_{1,2},\varphi_{1,3})U(\theta_2,\varphi_{2,2},\varphi_{2,3})\cdots U(\theta_k,\varphi_{k,2},\varphi_{k,3}) \cong U(\theta,\varphi_2,\varphi_3) \qquad (1.9)$$

That is the number of degrees of freedom after $k$ transformation remains three. However, four degrees of freedom are needed to realize the product states (1.6). Therefore, the states (1.6) cannot be realized. It follows that the products' states for $N \geq 2$ photons (1.4) also cannot be realized.

Remarks:

1. In fact, because (1.8) and (1.9) are the general representation of two dimensional unitary transformation (with $\det(u) = +1$ ), the conclusion holds to any unitary transformation.

2. The distinguishable states of photons may also be achieved via their polarization states rather then frequency. However, this doesn't change the above conclusion.

Next, we are going to see that non-unitary transformation allows us to construct the product space of $N$ distinguishable photons in two states.

## 4. The realization of the product states for two photons in two states

In Fig. 2 the non-unitary interferometer that can be used to produce realization of two photons in two states is described. The interferometer is constructed by:

1) a four beam splitter denoted as $i = \{A, B, C, D\}$ with corresponding transmitted and reflected coefficients $t_i, r_i$,
2) a four phase shifter, $\varphi_j$, $j = 1, 2, 3, 4$ denoted according to the arm they are located in
3) two detectors denoted $D_0^L, D_0^R$ are placed in the arms $|\ \rangle_{d_2}$ and $|\ \rangle_{c_1}$ respectively.

The subscript outside the ket indicates the arm the photons are in, for example $|1_{\omega^{(j)}}\rangle_{a_1}$ means that photon with frequency $\omega^{(j)}$ is in the arm $a_1$ and the subscript notation $|\ \rangle_0$ and $|\ \rangle_1$ are reserved to denote the original photon entering the beam splitter, i.e. before splitting.

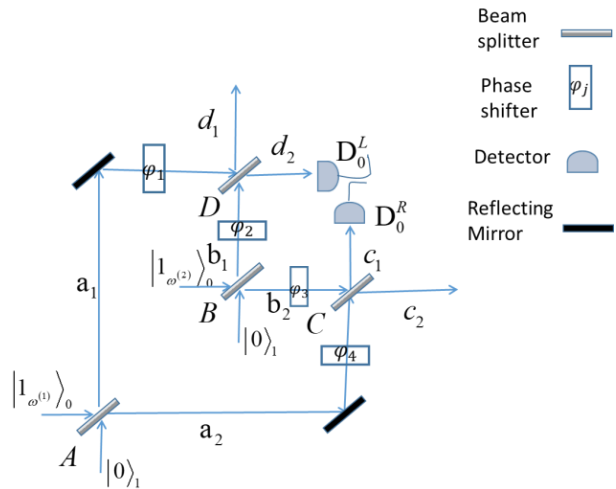

*Figure 2a non-unitary interferometer for the product states of two distinguished photons. See detail in the text*

The operation principle of the interferometer consists of taking two distinguishable photons one $|1_{\omega^{(1)}}\rangle$ of which has the frequency $\omega^{(1)}$ and enters into the beam splitter $A$, and the second photon $|1_{\omega^{(2)}}\rangle$ with frequency, $\omega^{(1)} \neq \omega^{(2)}$, that enters to the beam splitter $B$. The states of interest to us are the those at legs $d_1$ and $c_2$, provided that the detectors $D_L^0, D_R^0$ have negative readings, i.e. $D_L^0 = D_R^0 = 0$. Often this method of choosing states could be regarded as "postselected measurement" [5]. We will show now how the interferometer of figure 3 can be used to fully map the tensor product space of the two distinguishable photons in two states(1.6).

The photon states are given by $|1_{\omega^{(1)}}\rangle = \hat{a}^\dagger_{\omega^{(1)}}|0\rangle, |1_{\omega^{(2)}}\rangle = \hat{a}^\dagger_{\omega^{(2)}}|0\rangle$, where $|0\rangle$ is the vacuum state, $\hat{a}^\dagger_{\omega^{(1)}}, \hat{a}^\dagger_{\omega^{(2)}}$ and $\hat{a}_{\omega^{(1)}}, \hat{a}_{\omega^{(2)}}$ are the creation and annulation operators of the distinguishable photons with the frequencies $\omega^{(1)}$ and $\omega^{(2)}$.

With the following communication relation for the photons [3]

$$\left[\hat{a}^\dagger_{\omega^{(i)}}, \hat{a}^\dagger_{\omega^{(i')}}\right] = \left[\hat{a}_{\omega^{(i)}}, \hat{a}_{\omega^{(i')}}\right] = 0$$

$$\left[\hat{a}^\dagger_{\omega^{(i)}}, \hat{a}_{\omega^{(i')}}\right] = \left[\hat{a}^\dagger_{\omega^{(i)}}, \hat{a}_{\omega^{(i')}}\right] = \delta_{i,i'}$$

whereas $\{i, i'\} = 1, 2 \cdots N$ are the numbers that label the photons.

The wave function for the photon starts at A, $|1_{\omega^{(1)}}\rangle$ and travels throughout the interferometer is

$$A: |1_{\omega^{(1)}}\rangle_2 |0\rangle_1 \xrightarrow{A} e^{i\varphi_4} t_A \hat{a}^\dagger_{\omega^{(1)},a_2} + e^{i\varphi_1} r_A \hat{a}^\dagger_{\omega^{(1)},a_1} \begin{cases} \xrightarrow{C} e^{i\varphi_4} t_A r_C \hat{a}^\dagger_{\omega^{(1)},c_2} + e^{i\varphi_4} t_A t_C \hat{a}^\dagger_{\omega^{(1)},c_1} \\ \xrightarrow{D} e^{i\varphi_1} r_A r_D \hat{a}^\dagger_{\omega^{(1)},d_1} + e^{i\varphi_1} r_A t_D \hat{a}^\dagger_{\omega^{(1)},d_2} \end{cases}$$

and the wave function for the photon that starts at B, $|1_{\omega^{(2)}}\rangle$ and travels throughout the interferometer

$$B: |1_{\omega^{(2)}}\rangle_2 |0\rangle_1 \xrightarrow{B} e^{i\varphi_3} t_B \hat{a}^\dagger_{\omega^{(2)},b_2} + e^{i\varphi_2} r_B \hat{a}^\dagger_{\omega^{(2)},b_1} \begin{cases} \xrightarrow{C} e^{i\varphi_3} t_B r_C \hat{a}^\dagger_{\omega^{(2)},c_1} + e^{i\varphi_3} t_B t_C \hat{a}^\dagger_{\omega^{(2)},c_2} \\ \xrightarrow{D} e^{i\varphi_2} r_B t_D \hat{a}^\dagger_{\omega^{(2)},d_1} + e^{i\varphi_2} r_B r_D \hat{a}^\dagger_{\omega^{(2)},d_2} \end{cases} \quad (1.10)$$

If we select only the cases where detectors $D^R_0, D^L_0$ have zero readings, then the beam splitters $C(D)$ operate *as if* they were reflected mirrors, of the photons arriving from arm $|\ \rangle_{a_1(a_2)}$ and operate *as if* the total transmitted channels of the photon were arriving from arm $|\ \rangle_{b_1(a_2)}$.

That is,

a. the wave function of photon A, $|1_{\omega^{(1)}}\rangle$ on the arm $|1_{\omega^{(1)}}\rangle_{c_2}, |1_{\omega^{(1)}}\rangle_{d_2}$ becomes,

$$A: \to \begin{cases} e^{i\varphi_4} t_A r_C \hat{a}^\dagger_{\omega^{(1)},c_2} + e^{i\varphi_4} t_A t_C \hat{a}^\dagger_{\omega^{(1)},c_1} \xrightarrow{D^R_0 = 0} e^{i\varphi_4} t_A \hat{a}^\dagger_{\omega^{(1)},c_2} \\ e^{i\varphi_1} r_A r_D \hat{a}^\dagger_{\omega^{(1)},d_1} + e^{i\varphi_1} r_A t_D \hat{a}^\dagger_{\omega^{(1)},d_2} \xrightarrow{D^L_0 = 0} e^{i\varphi_1} r_A \hat{a}^\dagger_{\omega^{(1)},d_1} \end{cases}$$

b. the wave function of photon B, $|1_{\omega^{(2)}}\rangle$ on the arm $|1_{\omega^{(2)}}\rangle_{c_2}, |1_{\omega^{(2)}}\rangle_{d_2}$ becomes,

$$B:\to \begin{cases} e^{i\varphi_3}t_B r_C \hat{a}^\dagger_{\omega^{(2)},c_1} + e^{i\varphi_3}t_B t_C \hat{a}^\dagger_{\omega^{(2)},c_2} \xrightarrow{D_0^R=0} e^{i\varphi_3}t_B \hat{a}^\dagger_{\omega^{(2)},c_2} \\ e^{i\varphi_2}r_B t_D \hat{a}^\dagger_{\omega^{(2)},d_1} + e^{i\varphi_2}r_B r_D \hat{a}^\dagger_{\omega^{(2)},d_2} \xrightarrow{D_0^L=0} e^{i\varphi_2}r_B \hat{a}^\dagger_{\omega^{(2)},d_1} \end{cases}$$

Accordingly, the normalized wave function of the photons becomes

$$\left|1_{\omega^{(1)}}\right\rangle = \left(e^{i\varphi_4}t_A \hat{a}^\dagger_{\omega^{(1)},c_2} + e^{i\varphi_1}r_A \hat{a}^\dagger_{\omega^{(1)},d_1}\right)|0\rangle$$
$$\left|1_{\omega^{(2)}}\right\rangle = \left(e^{i\varphi_3}t_B \hat{a}^\dagger_{\omega^{(2)},c_2} + e^{i\varphi_2}r_B \hat{a}^\dagger_{\omega^{(2)},d_1}\right)|0\rangle$$

(1.11)

Since we have $|t_A|^2 + |r_A|^2 = |t_B|^2 + |r_B|^2 = 1$, the waves functions map as

$$\begin{aligned} e^{i\varphi_4}t_A &= \alpha_1^{(1)} & e^{i\varphi_1}r_A &= \alpha_2^{(1)} \\ e^{i\varphi_3}t_B &= \alpha_1^{(2)} & e^{i\varphi_2}r_B &= \alpha_2^{(2)} \end{aligned}$$

(1.12)

From (1.11) and (1.12) the tensor product states of the two distinguishable photons (1.6) are established.

Note that since a global phase is physically unimportant, we could use two phases instead of four. Nevertheless, using four phases somewhat simplifies the derivation. Now we can formulate the following effective *working principle* of the interferometer given in figure 2 as follows:

> At the output of the beam splitter $A$ and $B$ a general single photon in two states is prepared, a photon at arms $a_1, a_2$ and the second photon at arms $b_1, b_2$. Each of the photons are prepared according to the process that leads to realization of the state (1.5). Then, by the post- selected measurement of a negative result behind the arms $c_1, d_2$, i.e. $D_L^0 = D_R^0 = 0$, the photon states behind the beam splitter $C$ and $D$ are added. That is, the beam splitters $C(D)$ are effectively operating as total transmitters for photon in the incoming arms $b_2(b_1)$ and total reflectors for the photon in the incoming arms $a_2(a_1)$. This leads to product states of the two photons.

This effective *working principle* will be used below to simplify the discussion.

## 5. The realization of $N$ distinguishable photons in two states

In order to see the scheme of realization of the tensor product space for $N$ distinguishable photons in two states, let us construct first the general state of three distinguishable photons. To do that we will use the results from section 3 and use the effective working principle that is given there.

Denoting the general states of the third distinguishable photon by(1.5),

$$|1_{\omega^{(3)}}\rangle = \left(\alpha_1^{(3)} a^\dagger_{\omega^{(3)}} + \alpha_2^{(3)} a^\dagger_{\omega^{(3)}}\right)|0\rangle$$

the tensor product state of the three photons is

$$|1_{\omega^{(1)}}\rangle \otimes |1_{\omega^{(2)}}\rangle \otimes |1_{\omega^{(3)}}\rangle \quad (1.13)$$

First we prepare two of the photons, say $|1_{\omega^{(1)}}\rangle$ and $|1_{\omega^{(2)}}\rangle$, in their tensor product states via the interferometer as described in section 3. This reduces the states of three photons from the tensor product (1.13) to the tensor product

$$|1_{\omega^{(1)}}, 1_{\omega^{(2)}}\rangle \otimes |1_{\omega^{(3)}}\rangle \quad (1.14)$$

In figure 3 an interferometer of three photons is shown. This interferometer has been constructed such that the states at its internal legs $|\ \rangle_{d_1}$ and $|\ \rangle_{c_2}$ are the output states of the interferometer shown at figure 2, which is the tensor product state of two photons in two states.

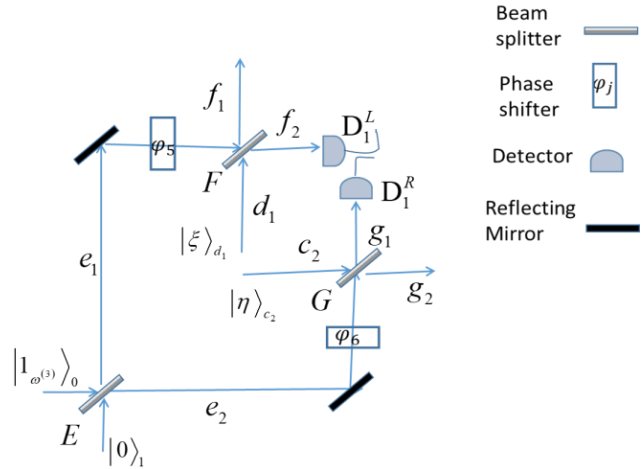

Figure 3 non-unitary interferometer for the produces stats of $N$ photons. See detail in the text

From Eqs (1.6), (1.11) and (1.12) the photons states at the internal legs of the interferometer at fig3 are,

$$\begin{aligned}|\xi\rangle_{d_1} &= \left(\alpha_1^{(1)} \hat{a}^\dagger_{\omega^{(1)}} + \alpha_1^{(2)} \hat{a}^\dagger_{\omega^{(2)}}\right)|0\rangle \\ |\eta\rangle_{c_2} &= \left(\alpha_2^{(1)} \hat{a}^\dagger_{\omega^{(1)}} + \alpha_2^{(2)} \hat{a}^\dagger_{\omega^{(2)}}\right)|0\rangle\end{aligned} \quad (1.15)$$

By (1.5) we have $\left|\alpha_1^{(1)}\right|^2 + \left|\alpha_2^{(1)}\right|^2 = \left|\alpha_1^{(2)}\right|^2 + \left|\alpha_2^{(2)}\right|^2 = 1$, thus the wave functions (1.15) are normalized.

The output states of $|\xi\rangle_{d_1}$ behind the beam splitter $F$ are

$$|\xi\rangle_{d_1} \xrightarrow{F} \begin{cases} t_f \left( \alpha_1^{(1)} \hat{a}^\dagger_{\omega^{(1)},f_1} + \alpha_2^{(2)} \hat{a}^\dagger_{\omega^{(2)},f_1} \right)|0\rangle \\ r_f \left( \alpha_1^{(1)} \hat{a}^\dagger_{\omega^{(1)},f_2} + \alpha_2^{(2)} \hat{a}^\dagger_{\omega^{(2)},f_2} \right)|0\rangle \end{cases} \quad (1.16)$$

and the output states of $|\eta\rangle$ behind the beam splitter $G$ are

$$|\eta\rangle_{c_2} \xrightarrow{G} \begin{cases} t_g \left( \alpha_2^{(1)} \hat{a}^\dagger_{\omega^{(1)},g_2} + \alpha_2^{(2)} \hat{a}^\dagger_{\omega^{(2)},g_2} \right)|0\rangle \\ r_g \left( \alpha_2^{(1)} \hat{a}^\dagger_{\omega^{(1)},g_1} + \alpha_2^{(2)} \hat{a}^\dagger_{\omega^{(2)},g_1} \right)|0\rangle \end{cases} \quad (1.17)$$

Now the post selected measurements with the detectors read no detection, $D_1^R = D_1^L = 0$ operate in the same way as in the case of two photons interferometer. Thus, using the effective working principle at section 3 the effectively transmitted and reflected coefficients for Eqs. (1.16), (1.17) become $t_{f/g}^{(efc)} \to 1$ and $r_{f/g}^{(efc)} \to 0$

That is, the post-selected states (1.16) and (1.17) are,

$$|\xi\rangle_{d_1} \xrightarrow{D_1^L = 0} \left( \alpha_1^{(1)} \hat{a}^\dagger_{\omega^{(1)},f_1} + \alpha_1^{(2)} \hat{a}^\dagger_{\omega^{(2)},f_1} \right)|0\rangle$$

$$|\eta\rangle_{c_2} \xrightarrow{D_1^L = 0} \left( \alpha_2^{(1)} \hat{a}^\dagger_{\omega^{(1)},g_2} + \alpha_2^{(2)} \hat{a}^\dagger_{\omega^{(2)},g_2} \right)|0\rangle$$

whereby using (1.5) are the normalized states.

Thus the product states of the photons $\omega^{(1)}$ and $\omega^{(2)}$ are

$$\left|1_{\omega^{(1)}}, 1_{\omega^{(2)}}\right\rangle = \left( \alpha_1^{(1)} \hat{a}^\dagger_{\omega^{(1)},f_1} + \alpha_2^{(1)} \hat{a}^\dagger_{\omega^{(1)},g_2} \right) \left( \alpha_1^{(2)} \hat{a}^\dagger_{\omega^{(2)},f_1} + \alpha_2^{(2)} \hat{a}^\dagger_{\omega^{(2)},g_2} \right)|0\rangle$$

Now, the state of the third photon $\left|1_{\omega^{(3)}}\right\rangle$ which enters into the beam splitter $E$ and passes through the phase shifters, $\varphi_5, \varphi_6$ . becomes

$$E: \left|1_{\omega^{(3)}}\right\rangle_2 |0\rangle_1 \xrightarrow{E} \left( e^{i\varphi_5} r_e a^\dagger_{\omega^{(3)},e_1} + e^{i\varphi_6} t_e a^\dagger_{\omega^{(3)},e_2} \right)|0\rangle$$

and then after passing the beam splitters $F$ and $G$ its states become

$$e^{i\varphi_6} r_e a^\dagger_{\omega^{(3)},e_2} |0\rangle \xrightarrow{F} \left( e^{i\varphi_5} r_e r_f a^\dagger_{\omega^{(3)},f_1} + e^{i\varphi_5} r_e t_f a^\dagger_{\omega^{(3)},f_2} \right)|0\rangle$$

$$e^{i\varphi_5} t_e a^\dagger_{\omega^{(3)},e_1} |0\rangle \xrightarrow{G} \left( e^{i\varphi_6} t_g t_e a^\dagger_{\omega^{(3)},g_1} + e^{i\varphi_6} r_g t_e a^\dagger_{\omega^{(3)},g_2} \right)|0\rangle$$

(1.18)

Repeating the above arguments for (1.18): Now the post selected measurements with the detectors read no detection, $D_1^R = D_1^L = 0$ operate in the same way as in the case of two photons interferometer. Thus, using the effective working principle at section 3 the effectively transmitted and reflected coefficients for Eqs.(1.18) become $t_{f/g}^{(\text{efc})} \to 1$ and $r_{f/g}^{(\text{efc})} \to 0$

Thus, the normalized post-selected photon state (1.18) is

$$|1_{\omega^{(3)}}\rangle = \left(e^{i\varphi_5} r_e a^\dagger_{\omega^{(3)}, f_1} + e^{i\varphi_6} t_e a^\dagger_{\omega^{(3)}, g_2}\right)|0\rangle$$

That is, by choosing the three parameters $\varphi_5, \varphi_6$, $r_E$ and the normalization condition $\left|\alpha_1^{(3)}\right|^2 + \left|\alpha_2^{(3)}\right|^2 = |r_E|^2 + |t_E|^2 = 1$ we have

$$e^{i\varphi_5} r_e = \alpha_1^{(3)}$$
$$e^{i\varphi_6} t_e = \alpha_2^{(3)}$$

Then, the final state of the three photons is the product states<

$$|1_{\omega^{(1)}}, 1_{\omega^{(2)}}, 1_{\omega^{(3)}}\rangle = |1_{\omega^{(1)}}, 1_{\omega^{(2)}}\rangle \otimes \left(\alpha_1^{(3)} a^\dagger_{\omega^{(3)}, f_1} + \alpha_2^{(3)} a^\dagger_{\omega^{(3)}, g_2}\right)|0\rangle$$

which is the realization of the tensor product space(1.14).

The general case of $N$ photons can be built by using the mathematical induction of the steps that have been followed for three photons above.

Accordingly, this can be shown as follows:

1. assuming that we can prepare the $N-1, N \geq 3$, photons in a general product states $|\xi\rangle_{d_1}$ and $|\eta\rangle_{c_2}$ as an input state of the interferometer shown in fig 3.

$$|\xi\rangle_{d_1} \otimes |\eta\rangle_{c_2} = |1_{\omega^{(1)}}\rangle \otimes |1_{\omega^{(2)}}\rangle \cdots \otimes |1_{\omega^{(N-1)}}\rangle$$

2. the $N$ th' photon enters into beam splitter $E$.
3. by choosing the value of the beam splitter, say $t_E$, and the two phases, $\varphi_5 \& \varphi_6$ the states of the photon can be chosen to be at one of any of the possible states of the photon.

$$|1_{\omega^{(N)}}\rangle = \left(t_E e^{i\varphi_5} a_{\omega^{(N)}, e_1} + r_E e^{i\varphi_6} a_{\omega^{(N)}, e_2}\right)|0\rangle$$

4. by applying the effective working principle at section 3, the post selected measurement for $D_1^L = D_1^R = 0$ gives:

4.1 the $N-1$ photons stats at $|\xi\rangle_{f_1}$ and $|\eta\rangle_{g_2}$ does not change.

$$|\eta\rangle_{f_1} \otimes |\xi\rangle_{g_2} = |1_{\omega^{(1)}}\rangle \otimes |1_{\omega^{(2)}}\rangle \cdots \otimes |1_{\omega^{(N-1)}}\rangle$$

4.2 for the $N$ th' photon the state at $|1_{\omega^{(N)}}\rangle_{f_1}$ and $|1_{\omega^{(N)}}\rangle_{g_2}$ is the same as at $|1_{\omega^{(N)}}\rangle_{e_1}$ and $|1_{\omega^{(N)}}\rangle_{e_2}$, therefore we have

$$|1_N\rangle = \left(e^{i\varphi_5} r_e a^\dagger_{\omega^{(N)}, f_1} + e^{i\varphi_6} t_e a^\dagger_{\omega^{(N)}, g_2}\right)|0\rangle$$

5. It follows that the state of $N$ photons at $|\ \rangle_{g_1}$ and $|\ \rangle_{f_2}$ is the tensor product of the $N-1$ photons at $|\ \rangle_{g_1}$ and $|\ \rangle_{f_2}$, with the $N$ th' photon state in these legs.

$$|\eta\rangle_{f_1} \otimes |\xi\rangle_{g_2} \otimes |1_{\omega^{(N)}}\rangle = |1_{\omega^{(1)}}\rangle \otimes |1_{\omega^{(2)}}\rangle \cdots \otimes |1_{\omega^{(N-1)}}\rangle \tag{1.19}$$

Eq (1.19) shows that the realization of the tensor product space of $N$ photons can be produced as claimed.

## 6. Summary and discussion

The aim of this paper was to show a way to construct a general realization of the tensor product space of $N$ distinguishable photons in two states. In section 2, we stated the aim of the paper: realization of the product states of $N$ photons in two states (1.4). In section 3 we have shown that such a realization cannot be done via a unitary transformation. In section 4 figure 2 a non-unitary interferometer that can be used for such a realization of two photons product space in two states was given. In section 5 a modified interferometer of figure 2 was given; see figure 3 that shows how to realize the product states of three photons. Then, based on this modified interferometer and via a mathematical induction, the generalization of the realization of the product state of $N$ photons in two states was shown. This realization from either perspective seems to be new.

The results of the paper may be understood from two different perspectives. First, the fundamental ability to give a realized map of the photons product space in two states. Then, in view of the so called "Second Quantum Revolution" it is a technical challenge to construct such a realization.

In this paper we focus on $N$ distinguishable photons in two states. However, the same interferometers (fig 2 and fig 3) can be used for identical photons to build the product states of $N$ indistinguishable photons. This possibility will be explored separately.